\documentclass[preprint,12pt]{elsarticle}

\usepackage{graphics}

\usepackage{amssymb}

\usepackage{lineno}
\usepackage{url}

 \biboptions{comma,square}

\begin{document}

\begin{frontmatter}



\title{Development of an anti-Compton veto for HPGe detectors operated in liquid argon using Silicon Photo-Multipliers}




\author{\corref{cor1}J. Janicsk\'o Cs\'athy}\ead{janicsko@mppmu.mpg.de}
\author{H. Aghaei Khozani} 
\author{A. Caldwell}
\author{\fnref{label1}X. Liu}
\author{B. Majorovits }

\address{Max-Planck-Institut f{\"ur} Physik, F\"oringer Ring 6, M\"unchen 80805, Germany}
\cortext[cor1]{Corresponding author:}
\fntext[label1]{Present Address: Shanghai Jiaotong University
Dongchuan Road 800,
Shanghai, P.R.China 200240
and
Center for High Energy Physics
Peking University
No.5 Yiheyuan Road Haidian District,
Beijing, P.R.China 100871}

\begin{abstract}

A proof of concept detector is presented for scintillation light detection in 
liquid argon using Silicon Photo-Multipliers. The aim of the work is to build an anti-Compton 
veto for germanium detectors operated directly in liquid argon as in the GERDA experiment. 
Wavelength shifting fibers are used to collect the scintillation light and to guide it to 
Multi-Pixel Photon Counters (MPPC).
Sufficient light yield was achieved to realize an effective anti-Compton veto.
Properties of the MPPC were studied at cryogenic temperatures and are additionally reported.

\end{abstract}


\end{frontmatter}


\section{Introduction}
\label{intro}

One of the biggest challenges in dark matter and double beta decay  experiments is to achieve 
a background level that is smaller than the expected signal. 
A reduction of the background can be achieved by adding shielding to the experiment and by careful 
selection of the components. 
After the limit imposed by the radioactivity of the cleanest materials available has been reached 
further reduction is possible only with an active veto system.

The GERDA \cite{GERDA}, \cite{GERDA2} experiment currently being commissioned 
is built for the search for neutrino-less double beta decay ($0\nu2\beta$) with High Purity Germanium detectors (HPGe).
The HPGe detectors are submerged directly in liquid argon (LAr). 
The LAr simultaneously acts as passive shielding and cryogenic cooling liquid.

The main source of background for the $0\nu2\beta$ decay is the radioactivity of the surrounding material.
$\gamma$ photons hitting the HPGe detector undergo Compton scattering and very often deposit only part of 
their energy before escaping the detector.
The half life limit on $0\nu2\beta$ decay at the end will be determined mainly by the number of these kind of events.

The escaping $\gamma$ photons with high probability will deposit their energy in the nearby environment. 
If the nearby space is filled with scintillating material the escaping  $\gamma$ photons can be detected 
and the events with partial energy deposition in the HPGe detector can be vetoed. 
Such an arrangement is called an anti-Compton veto because it strongly suppresses the Compton background 
between the $\gamma$ lines.

In GERDA the liquid argon was chosen primarily because of its higher density compared to liquid nitrogen.
In addition to that LAr, like all noble liquids, has the advantage of being an excellent scintillator. 
Ar scintillates in the VUV range emitting light at 128 nm. The properties of the scintillation light are well 
understood and described in the literature (see for example \cite{Hitachi}). 
The scintillator properties of LAr offer the possibility to use the scintillation light as an anti-Compton veto for the HPGe detectors, 
a possibility which is unexploited so far in the current design of GERDA.

Earlier it was shown that with a LAr anti-Compton veto
more than an order of magnitude suppression of the background can be achieved \cite{LArGe} around the Q value of the $0\nu2\beta$ decay. 
Such a huge improvement could be decisive for the success of GERDA if it could 
be realized without increasing the radioactive background or changing the original design too much.

Traditionally an anti-Compton veto like in \cite{Beetz1977353} is built using Photo-Multiplier Tubes (PMTs). 
The setup in \cite{LArGe} was also realized with an 8" PMT submerged in LAr.

Despite their widespread use PMTs have some important disadvantages. 
They are not well suited for a cryogenic environment and they 
contain components such as glass which have significant radioactive impurities. Furthermore the use of high voltage in 
an argon atmosphere will very often lead to spark breakdown. 

In the field of low background experiments considerable efforts are invested in finding an alternative of PMTs. 
Silicon Photo-Multipliers (SiPMs) are considered for experiments like NEXT \cite{Diaz} and they were investigated to be used in the XENON experiment
\cite{Aprile2006215} as well. 

In the experiment presented in this paper we investigated the possibility to replace the PMT with SiPMs in a setup 
similar to \cite{LArGe}.
SiPMs do not require high voltage and they have a quantum efficiency at least as high as PMTs.
Their small mass limits their contribution to the radioactive background. 

A major disadvantage of existing SiPMs is their small active area.  
We will try to overcome this disadvantage by connecting  wavelength shifting (WLS) fibers to it.
The WLS fiber collects light on a much bigger surface and guides the trapped photons to the SiPM.
Such a design would have the advantage of being compatible with the string design of GERDA.

\begin{figure}[h!]
\centering 
\includegraphics[width=0.40\columnwidth]{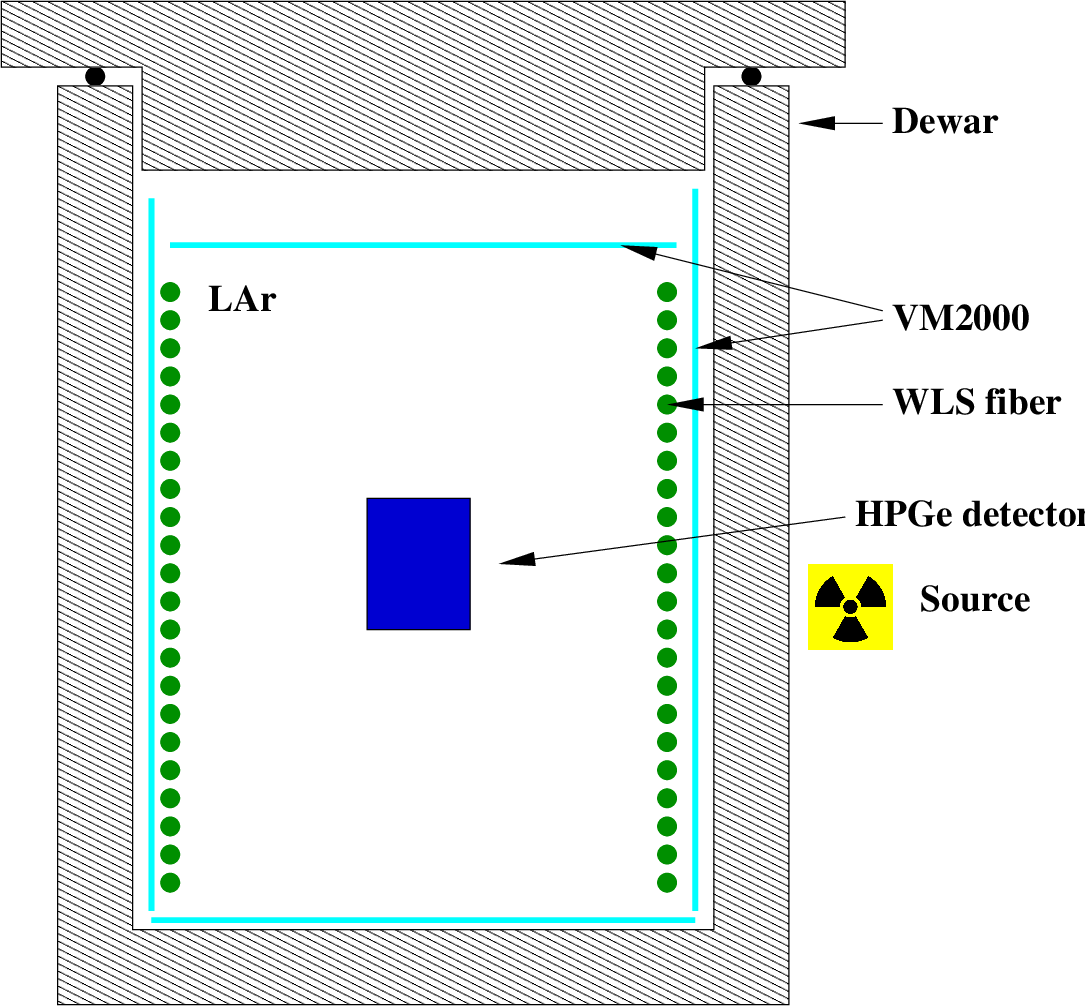}
\caption{\label{fig:wls_setup_1} 
Conceptual drawing of the tested setup.
Wavelength shifting fibers are placed in a volume of liquid argon delimited by reflector foils. 
The WLS fibers are connected to SiPMs (not shown). The HPGe detector is placed in the middle of the volume.}
\end{figure}

To test the principle a small scale experiment was set up to measure the properties of SiPMs 
in a cryogenic environment and to demonstrate a reasonably high light collection efficiency in LAr.
A small (20 l) dewar filled with LAr was instrumented with SiPMs and WLS fibers such that a HPGe detector 
could also be inserted in the active volume (see Fig.\ref{fig:wls_setup_1}).  

An important advantage of such a setup would be the reduced radioactivity.
We estimate that the integrated radioactivity of the materials used 
in our detector is about two orders of magnitude smaller than a single low background 8" PMT.
The activity of the WLS fiber was measured with the GeMPI detector at the Gran Sasso underground laboratory \cite{Heusser2006495}.
After measuring 20 m (22 g) of WLS fiber for one month, in the absence of a positive signal 
the following upper limits were obtained: 
$^{226}$Ra(U) $<$ 3.9 10$^{-9}$g/g, $^{228}$Th(Th) $<$ 1.3$\cdot$10$^{-9}$ g/g and $^{40}$K $<$ 4.6$\cdot$10$^{-6}$ g/g.
The measured limits are about an order of magnitude smaller than the activity of the low activity glass used for PMTs 
(see for example \cite{Arpesella20021}) and the weight of the components is another one order of magnitude smaller.  

A further advantage is the less bulky design and the lower weight of the components which simplifies the integration into an experimental setup. 

The main goal of this work is to provide a light detection solution for low background liquid scintillator 
experiments in general that is compatible with the stringent radiopurity requirements of dark matter and double beta decay experiments. 

Before describing the performance of our complete setup, we describe a series of measurements 
which were used to set the operating parameters of the SiPMs.

\section{Properties of SiPMs in a cryogenic environment}

The Silicon Photo-Multiplier is a multipixel Geiger-mode avalanche photodiode. 
It promises single photon resolution and good photon detection efficiency. 
For a detailed description and possible applications see for example \cite{Dolgoshein}.

The SiPM we chose for our experiment was  the Multi-Pixel Photon Counter (MPPC) made by Hamamatsu with 1 mm$^2$ active surface. 
We investigated the versions in ceramic case S10362-11-025C / -050C /-100C with 1600, 400 and 100 pixels respectively.

MPPCs were already characterized at liquid nitrogen (LN) temperatures in \cite{Otono} and \cite{Akiba}.
We have studied the temperature dependence of the dark rate and the break-down voltage of the specimens we used.
We also describe in this section the electronic setup that was used during the experiment described
 in Sec.\ref{section3} and Sec.\ref{section4}.

\begin{figure}[h!]
\includegraphics[width=0.5\columnwidth]{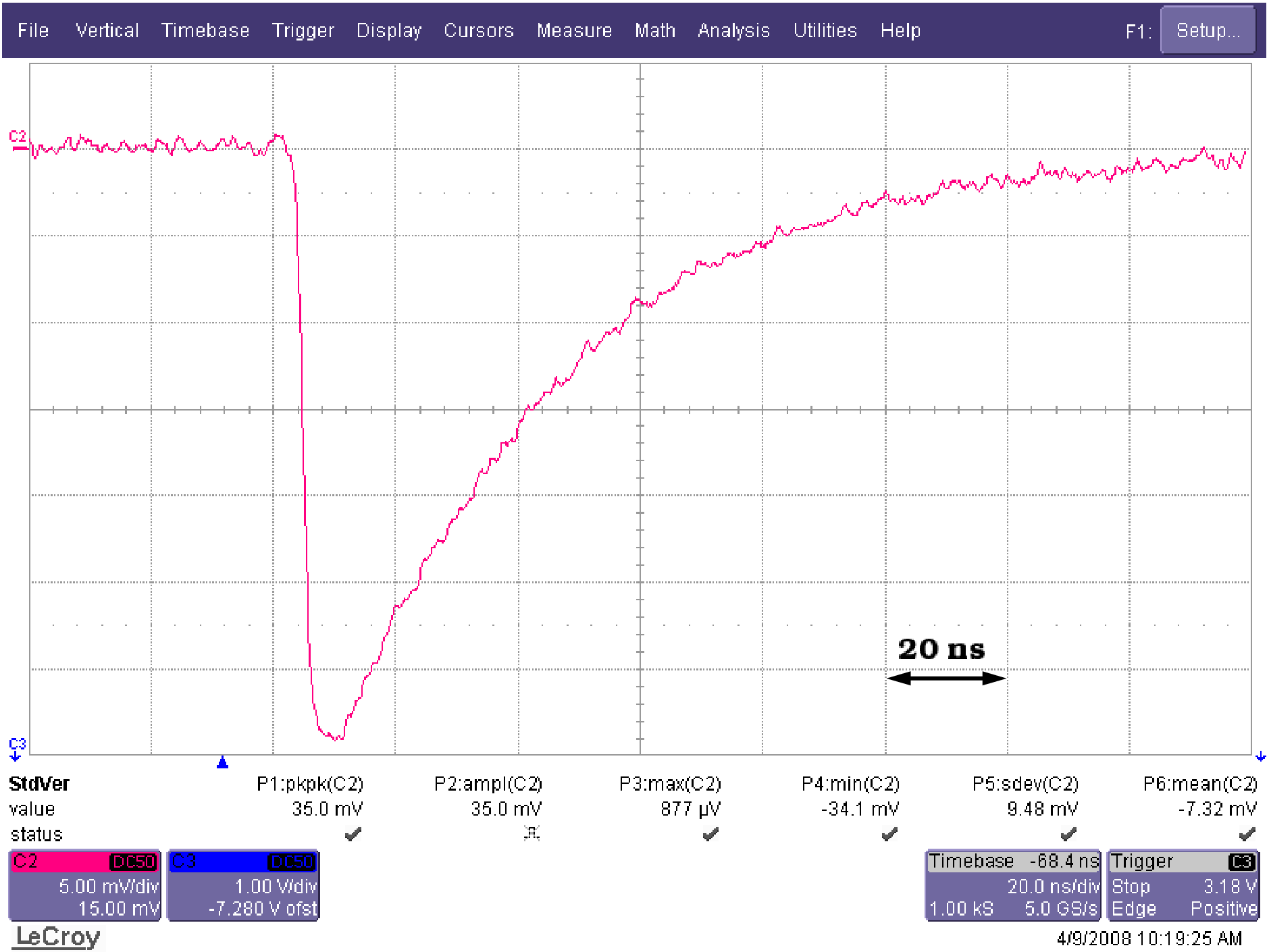}\includegraphics[width=0.5\columnwidth]{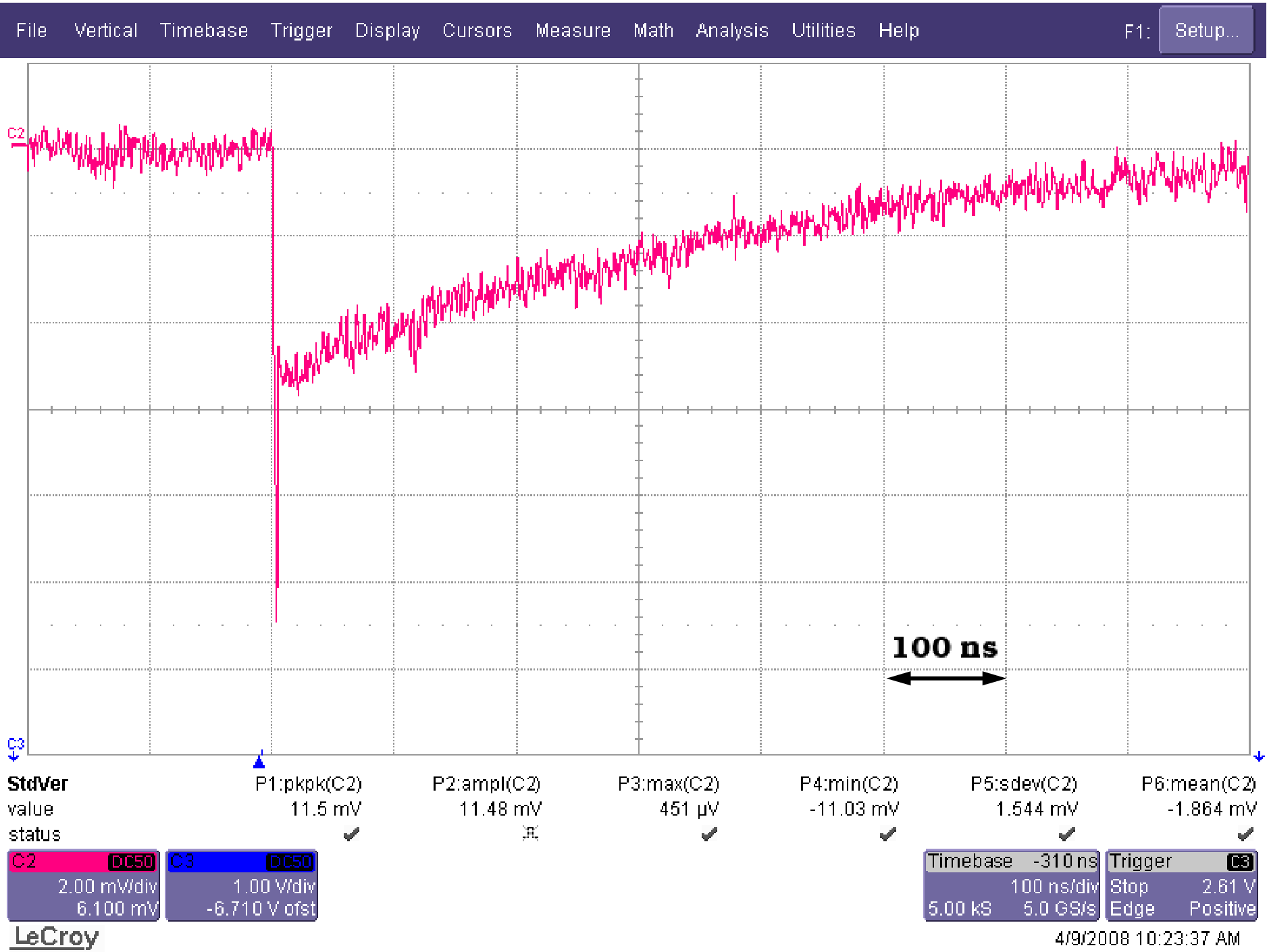}
\caption{Difference in pulse shape at room temperature and in liquid nitrogen. 
The waveforms were recorded with an oscilloscope using a 100 pixel MPPC.}
\label{fig:pulse_shape_LN}
\end{figure}

One property of the MPPCs is that the pulse shapes change at low temperatures.
As it was reported in \cite{Otono} the MPPC pulses at low temperatures start with a sharp peak and end with 
a decay time much longer than at room temperature (Fig.\ref{fig:pulse_shape_LN}). 
The change is related to the temperature dependence of the  quenching resistor. 
The mechanism responsible for the pulse shape is explained in \cite{Otono2}.

The consequence of the longer decay times is that the amplitude of the pulses is 4-5 times smaller in LN while 
the total charge contained in a one pixel pulse is the same as at room temperature at the same over-voltage.
For this reason and because the dark rate becomes negligible at LN temperature it was a natural choice to use charge sensitive amplifiers. 
The use of charge sensitive amplifiers makes our measurement of the gain of the SiPM independent of the temperature 
and it also helps to improve the signal to noise ratio.



\begin{figure}[h!]
\centering
\includegraphics[width=0.6\columnwidth]{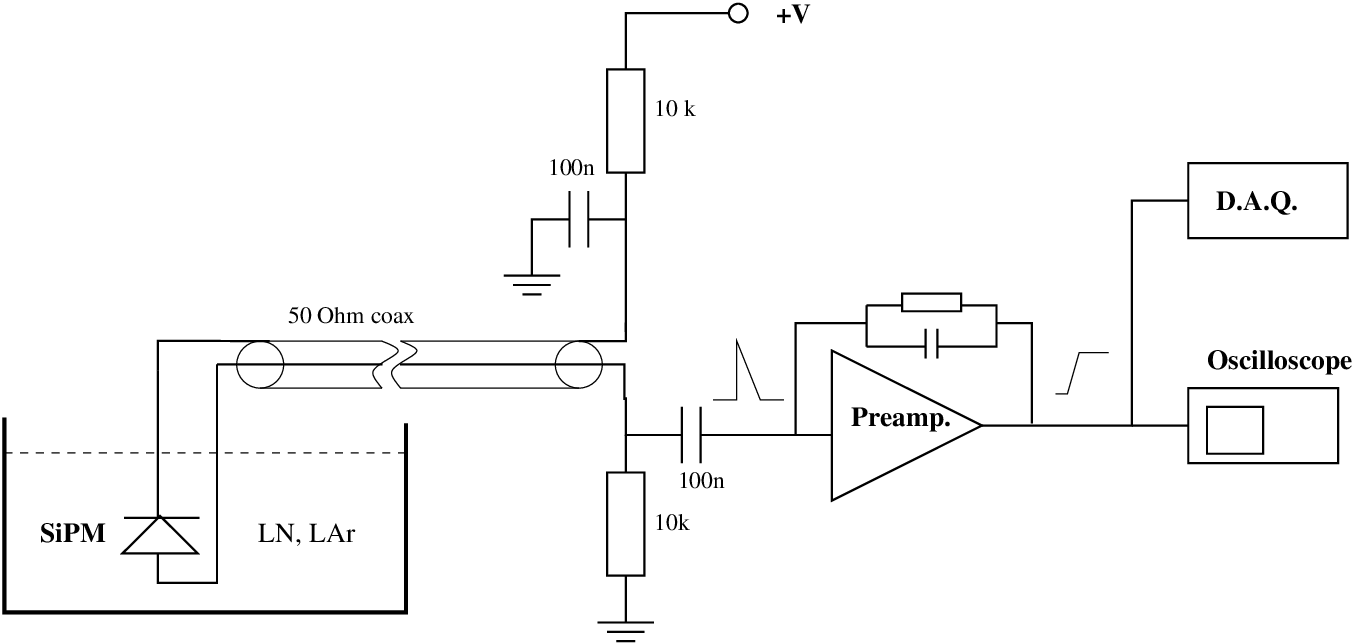}\\
\caption{\label{fig:e_schema} Schematic drawing of the electronic readout.
The same setup was used for the electronic measurements and for the scintillation light detection in Sec.\ref{section3}.
To record the pulse shapes in Fig.\ref{fig:pulse_shape_LN} the same setup with a linear amplifier was used.}
\end{figure}

\begin{figure}
\centering
\includegraphics[width=0.8\columnwidth]{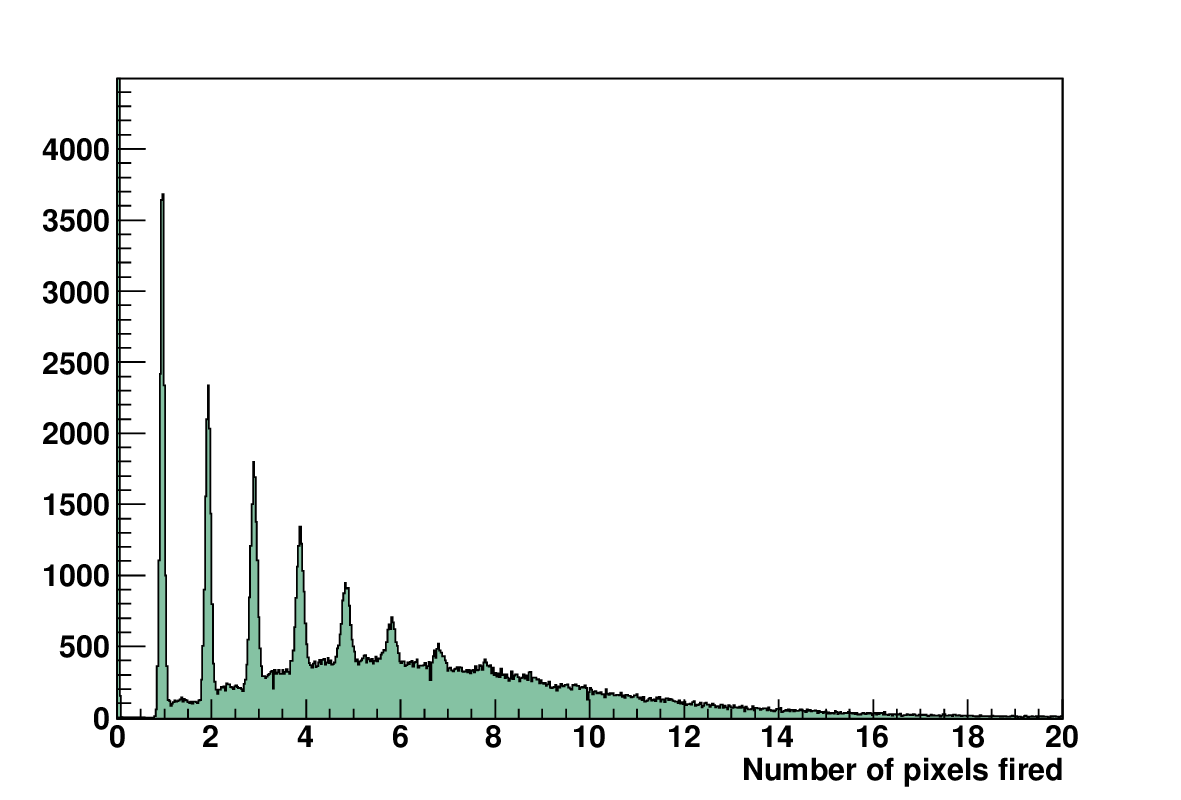}
\caption{\label{fig:SiPM_spctrm} Pulse height spectrum recorded with a 400 pixel MPPC:
the MPPC is separated from the preamplifier by a 50 cm cable and is submerged in LN.
}
\end{figure}

It was found that the CREMAT CR-112 \cite{CREMAT} amplifier has the optimal gain to match the input range of our DAQ.
The preamplifiers were operated at room temperature. The schematic drawing of the electronic 
read-out chain is shown in Fig.\ref{fig:e_schema}.
The MPPC was separated from the preamplifier by a 50 cm long cable. 
The integrated pulses were recorded with an XIA Pixie-4 DAQ \cite{XIA} with 75 MHz sampling rate. 
To measure the amplitude of the pulses we used the built in trapezoidal energy filter of the DAQ. 
The results were cross checked with off-line amplitude measurement on the recorded wave-forms.
In Fig.\ref{fig:SiPM_spctrm} the pulse height spectrum of a 400 pixel MPPC operated in liquid nitrogen is shown.

To measure the dark rate at different temperatures we wrapped the MPPCs in several layers of black tape and suspended them inside a 
dewar filled with LN. In the same experiment we measured three specimens with 100, 400 and 1600 pixels.
A Pt-100 temperature sensor was also attached to the SiPMs. As the LN slowly evaporated we recorded dark pulses with the DAQ.
At temperatures below 150 K the data acquisition was running for several hours until the peak corresponding to a single fired pixel could be seen.

During one measurement the temperature was stable within a few degrees. 
For each temperature we recorded data at several bias voltages. 
From the amplitude of the peak corresponding to a single fired pixel we could estimate the gain. 
From the bias voltage dependence of the gain we obtained the break-down voltage for each temperature by extrapolating the linear fits to zero gain.  
With this information we could calculate the optimal bias voltage for LAr temperature for the measurement described in Sec.\ref{section3} 

\begin{figure}
\centering
\includegraphics[width=0.8\columnwidth]{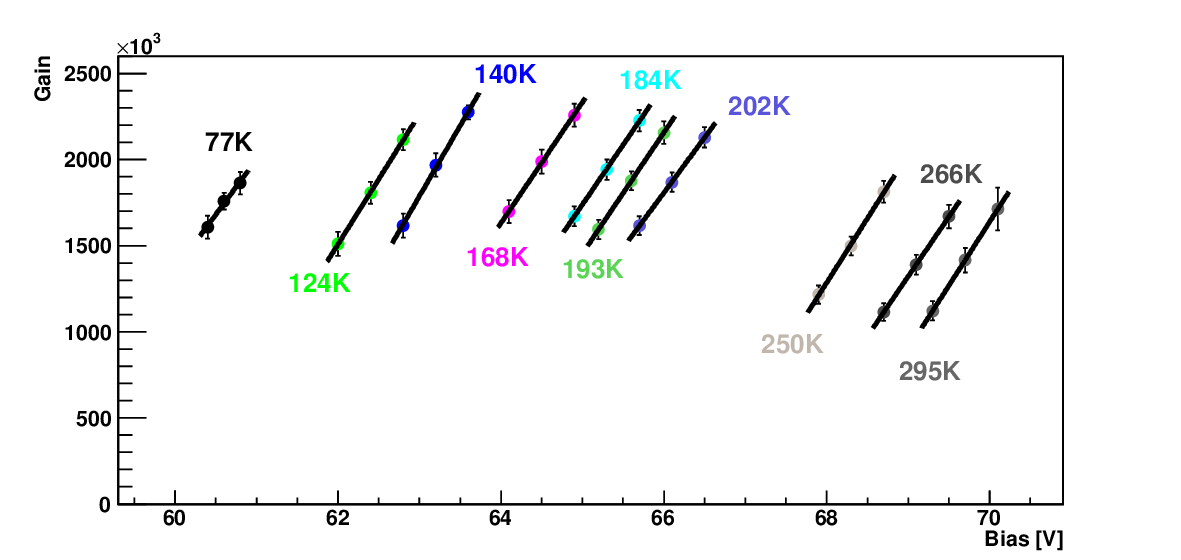}
\caption{\label{fig:SiPM_gain} Gain as a function of bias voltage at different temperatures measured with the 400 pixel MPPC.}
\end{figure}

Since in the final experiment we used only the 400 pixel MPPC 
here we show only the plots recorded with this device. The plots for the 100 and 1600 pixel devices look very similar.
Fig.\ref{fig:SiPM_gain} shows gain measurements at different temperatures and bias voltages. 

\begin{figure}
\centering
\includegraphics[width=0.8\columnwidth]{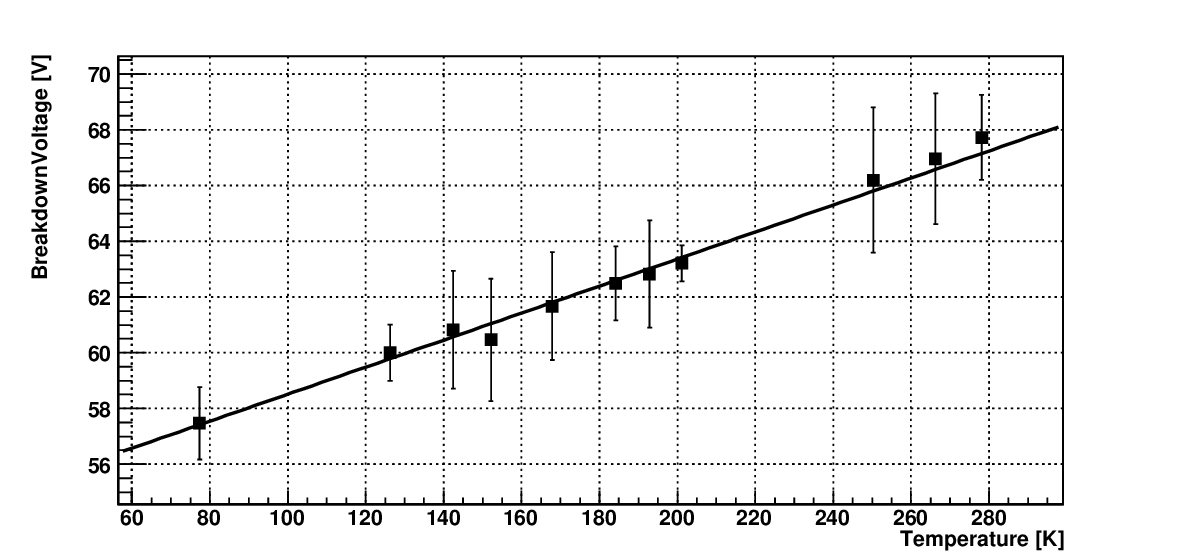}
\caption{\label{fig:SiPM_BDV} Break-down voltage as a function of temperature of the MPPC with 400 pixels.}
\end{figure}

\begin{figure}
\centering
\includegraphics[width=0.8\columnwidth]{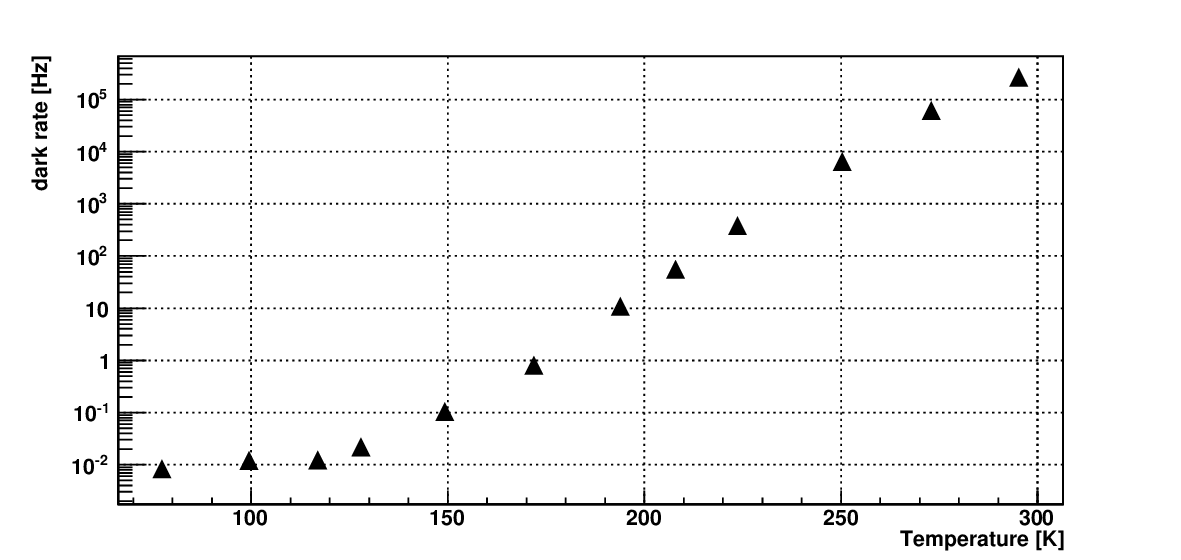}
\caption{\label{fig:SiPM_DR} Dark rate of the 400 pixel MPPC as a function of temperature at 2.5 V over-voltage. 
The error bars from the statistical error are smaller than the symbols.}
\end{figure}

After the temperature dependence of the breakdown voltage was known (Fig.\ref{fig:SiPM_BDV}) 
we remeasured the dark rate as a function of temperature always at the same over-voltage. The results are shown in Fig.\ref{fig:SiPM_DR}.
The measured dark rate at LN temperature is at the level of 10$^{-2}$ Hz
at an over-voltage of 2.5 V for the 400 pixel device. This is four orders of magnitudes smaller than what is reported in \cite{Dolgoshein}
but agrees rather well with \cite{Akiba}. 
For the MPPCs with 100 and 1600 pixels the dark rate at liquid nitrogen temperature is within the same order of magnitude as for the 400 pixel MPPC.

\clearpage

\section{Experimental setup}
\label{section3}

Attempts to use WLS fibers in combination with liquid scintillators were already made before. The 
description of a similar setup to ours can be found in \cite{McKinsey}.

The detection of the scintillation light from argon presents a major challenge.
The 128 nm light emitted by LAr cannot be absorbed by a WLS fiber or any plastic lightguide since
such a short wavelength would be absorbed already in the cladding of the fiber before reaching the scintillating core.
We therefore coated a reflector foil (VM2000) with TPB (tetra-phenyl-butadiene) to serve as a primary wavelength shifter.
The LAr volume delimited by the mirror foil creates an optical cavity where the only absorber is the WLS fiber. 
This arrangement increases substantially the amount of light collected in the fiber.

\subsection{Choice of the WLS fiber}
The emission spectrum of the TPB \cite{TPB} and the absorption spectrum of most WLS fibers partially overlap. 
We chose the BCF-91A made by Saint-Gobain \cite{bicron} which has a wide absorption peak. 
Using the spectra from \cite{TPB} and \cite{bicron} we estimated the conversion efficiency from 128 nm to 500 nm to be around 50\%.  

The choice of the WLS fiber was also motivated by the typical size of available fibers. 
The 1 mm $\times$ 1 mm  square fiber is a reasonable match for the 1 mm$^2$ active area of the SiPM.

The major limiting factor in such a setup is the low trapping efficiency of the WLS fibers. 
The fibers with the highest trapping efficiency are the multiclad fibers with square cross section. 
The fiber used in the experiment had a trapping efficiency of 7.3\% and an attenuation length of $>$3.5 m \cite{bicron}. 

\subsection{Optical coupling}
\begin{figure}[h!]
\centering 
\includegraphics[width=0.30\columnwidth]{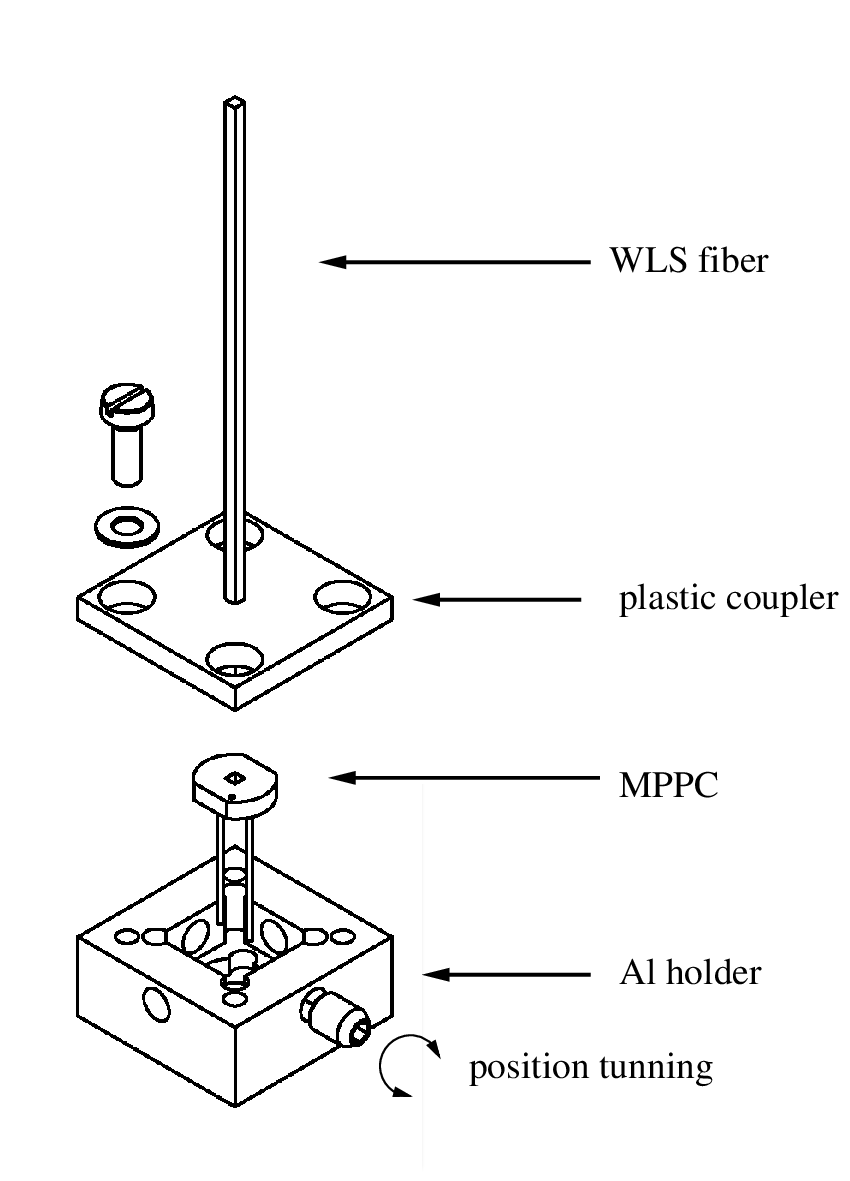}\\
\caption{\label{fig:coupling} Optical coupling between the WLS fiber and the SiPM.
 The dimensions of the assembled coupling are 16 mm x 16 mm x 9 mm.} 
\end{figure}

The optical connection between the fiber and the SiPM was realized as follows: 
First the fiber was glued in a squared shape plastic connector (Fig.\ref{fig:coupling}) taking care to keep the edges 
of the square fiber aligned with the connector. The fiber end was polished together with the connector.
The SiPM was placed in an Al holder that was also designed to keep the SiPM aligned and prevent twisting.  
The plastic connector was fixed with screws to the Al holder. 
With the screws on each facet of the holder we could shift the plastic 
connector by $\pm$1 mm in all directions in the plane of the SiPM.  
The relative position of the fiber to the SiPM was tuned in the dark room using a LED pulser and monitoring 
the signal of the SiPM with an oscilloscope. In the position where the SiPM signal was maximal, 
the plastic connector was fixed. The screws fixing the connector in the final position were also pressing 
it down on the SiPM to ensure a close contact. 

The sensitive surface of the MPPC is 
protected by a 0.3 mm thick epoxy resin that prevents direct contact with the polished end of the fiber. 
From the total reflection angle inside the fiber we can calculate that the total illuminated 
area is about 35\% larger than the surface of the MPPC. 
For this reason we assume that the optical coupling causes a further 35\% loss of light at most.

\subsection{Mechanical setup}
\begin{figure}[h!]
\centering 
\includegraphics[width=0.50\columnwidth]{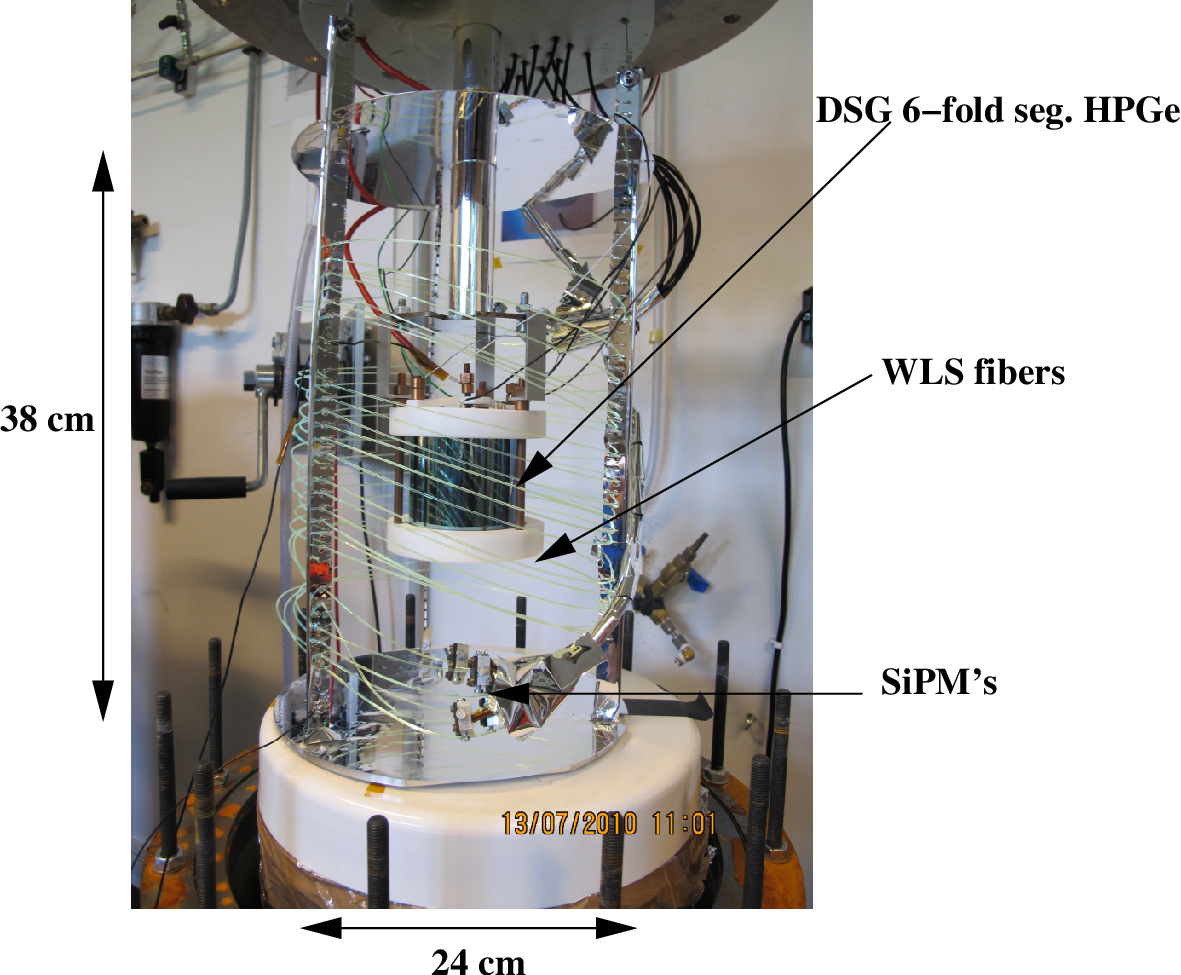}\\
\caption{\label{fig:wls_setup} 
Practical realization of the setup shown in Fig.\ref{fig:wls_setup_1}.
An Al frame holds six 2.5m WLS fibers and 12 SiPMs inside the optical couplers. 
In the middle is the HPGe detector which was inserted only for the last experiment. 
Before the experiment the whole structure is lowered in the dewar below.
The dewar has its walls covered with TPB coated VM2000 foil.}
\end{figure}

To hold the fibers inside the dewar a simple aluminium structure was designed. 
Three vertical bars with holes in them keep the coil of WLS fiber 
in place about 1 cm distance from the wall of the dewar. 
The distance between two parallel fibers is 1 cm. 
The holder with the fibers is shown in Fig.\ref{fig:wls_setup}. 
The volume enclosed by the fiber coil is approximately 17 l (23.8 kg Ar). 
The VM2000 foil coated with a TPB solution was glued to the internal walls of the dewar and  
the radioactive sources were placed immediately next to the dewar.

The thickness of the LAr between the HPGe detector and the vertical walls of the dewar is only about 7 cm. 

The dewar was gas tight and it was kept under small over pressure. To remove the air 
it was flushed with Ar gas for about an hour before the filling with LAr started.

\subsection{Light yield estimation}
\label{Yield_estimate}

Our light yield prediction is based on assumptions and estimates when measurements are not available.
One important assumption we make is that the light inside our LAr volume is distributed uniformly across the fibers. 
In this case the light collected at the end of the fiber will be:
 
\begin{equation}
I=I_{0}\frac{1}{L}\int_{0}^{L}e^{-\frac{x}{\lambda}}dx=I_{0}\frac{\lambda}{L}(1-e^{-\frac{L}{\lambda}})\label{eq:1}\end{equation}

where $I_{0}$ is the light intensity absorbed in the fiber, $\lambda$ is the attenuation length and L is the length of the fiber.

Another important property of the setup is that the fibers are inside an optical cavity and the light that is not absorbed 
in the fiber will be reflected back and forth many times. The intensity that is finally absorbed in the fiber will be proportional 
to $\frac{S}{1-(1-S)R}$, where $S$ is the ratio of the fiber surface to the reflector surface and $R$=0.95 is the reflectivity of the VM2000 
foil taken from \cite{vm2000}.

Following \cite{McKinsey} the photo electron (p.e.) yield can be calculated as;

\begin{equation}
N=Y E_{F} E_{tr}I\frac{S}{1-(1-S)R} \times PDE\label{eq:3}\end{equation}

where Y = 40000 is the number of primary photons per MeV deposited energy, $E_{tr}$ = 0.073 is the trapping efficiency of the square multiclad fiber \cite{bicron}, 
PDE = Photon Detection Efficiency of the MPPC, taken to be 0.25 at 500 nm (see \cite{Eckert}) and  
$E_{F}$ = fluor efficiency, the mean number of 500 nm photons produced by a 128 nm photon. 
We assume that $E_{F}$ is limited only by the absorption spectrum of the fiber (0.5 assumed). 

\begin{figure}[h!]
\centering
\includegraphics[width=0.8\columnwidth]{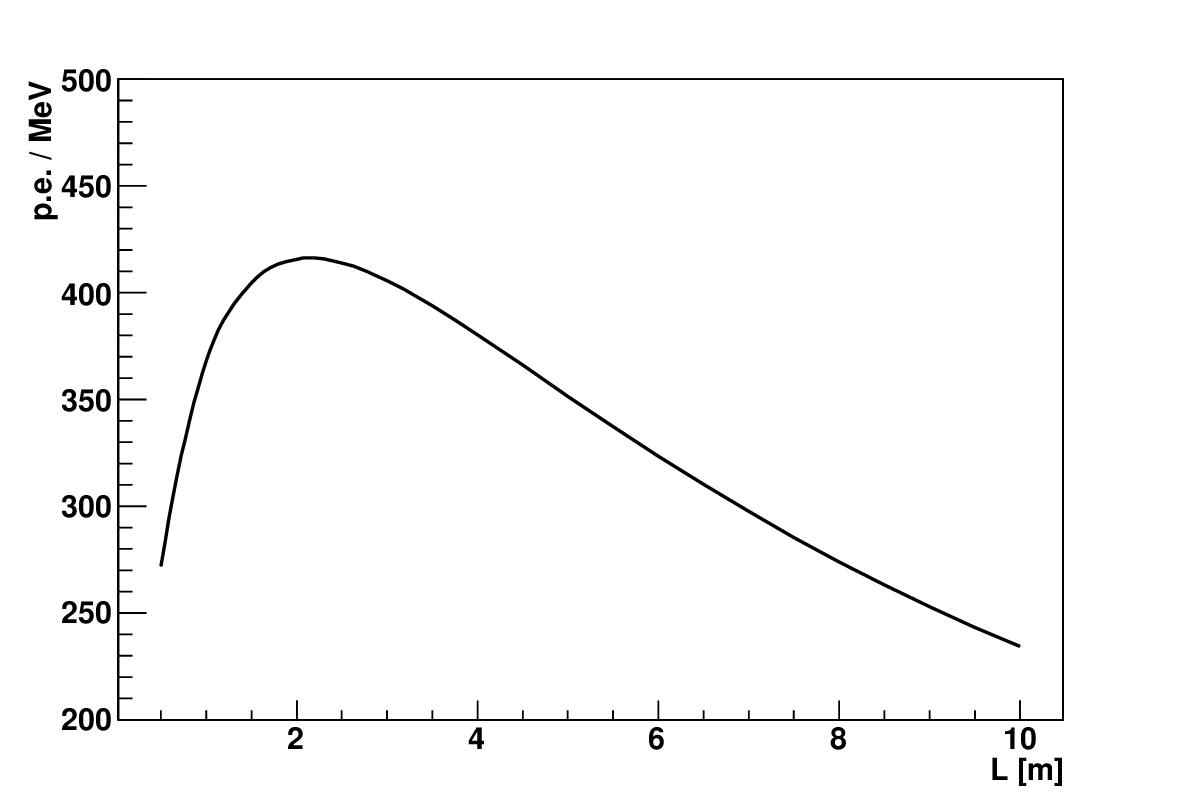}\\
\caption{\label{fig:yield_est} Estimated light yield as a function of fiber length for a six 1 mm $\times$ 1 mm fiber setup in a 17 l volume optical cavity.}
\end{figure}

The formula was evaluated for the geometry of our setup. The light yield will depend on the number of fibers used for the same total length.
Because of the limitations imposed by the number of feedthroughs and electronic channels we used six fibers with a SiPM on each end. 
In the estimated light yield as a function of fiber length (Fig.\ref{fig:yield_est}) there is a clear maximum between 2 and 3 meters. 
For this reason we chose to use 2.5 m long fibers. 
Based on this rough estimation in the experiment we expected to see about 400 p.e./MeV.

\subsection{Measured light yield}

Once the dewar was filled with liquid argon the setup was illuminated with $^{60}$Co and $^{228}$Th sources located just outside the dewar.
$^{60}$Co emits two $\gamma$ photons in coincidence at 1173.2 keV and 1332.5 keV. 
$^{228}$Th has a strong line at 2.6 MeV. 

In the recorded spectra we expect to see an enhancement corresponding to about 1.2 MeV and another to 2.6 MeV respectively.
 
An event was recorded by the DAQ when at least two SiPMs gave a signal above a 1 p.e. threshold. 
For each event we recorded 6 $\mu$s long pulse shapes which were used for off-line amplitude measurement.

All 12 channels were calibrated separately and a sum spectrum of total number of SiPM pixels fired was created.

\begin{figure}[h!]
\centering
\includegraphics[width=0.5\columnwidth]{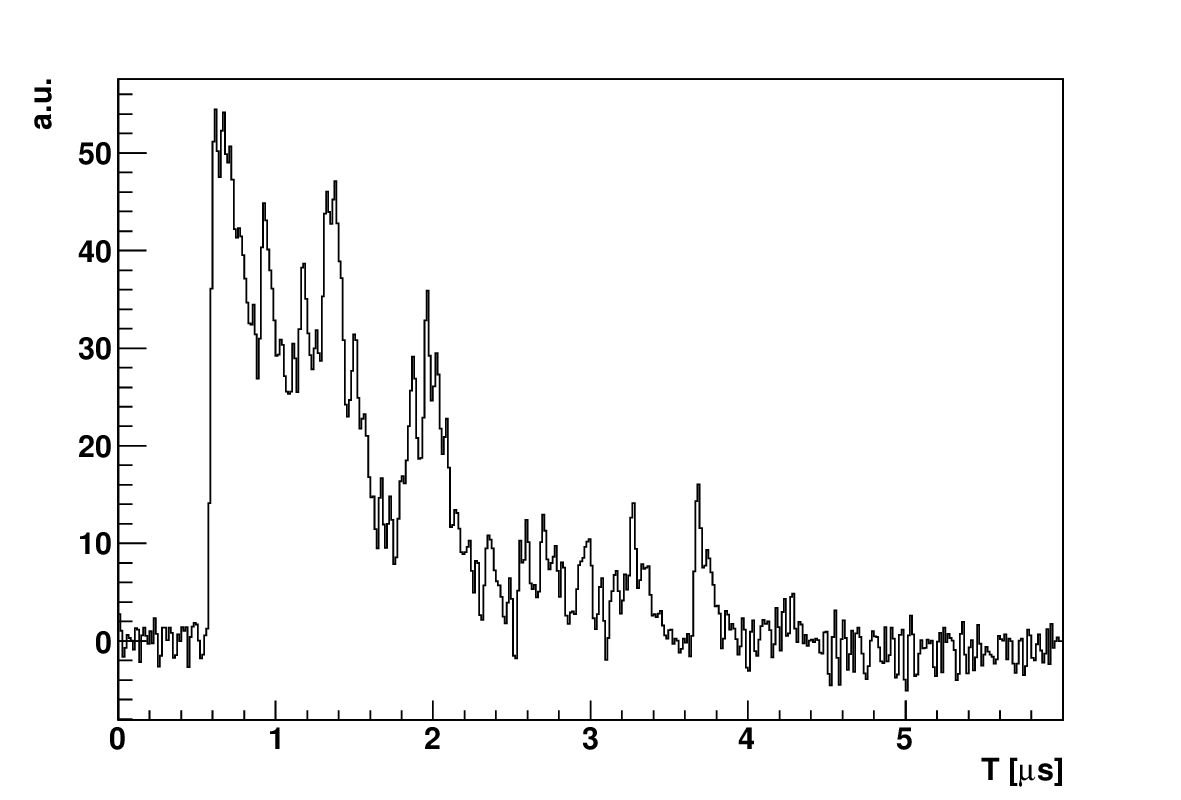}\includegraphics[width=0.5\columnwidth]{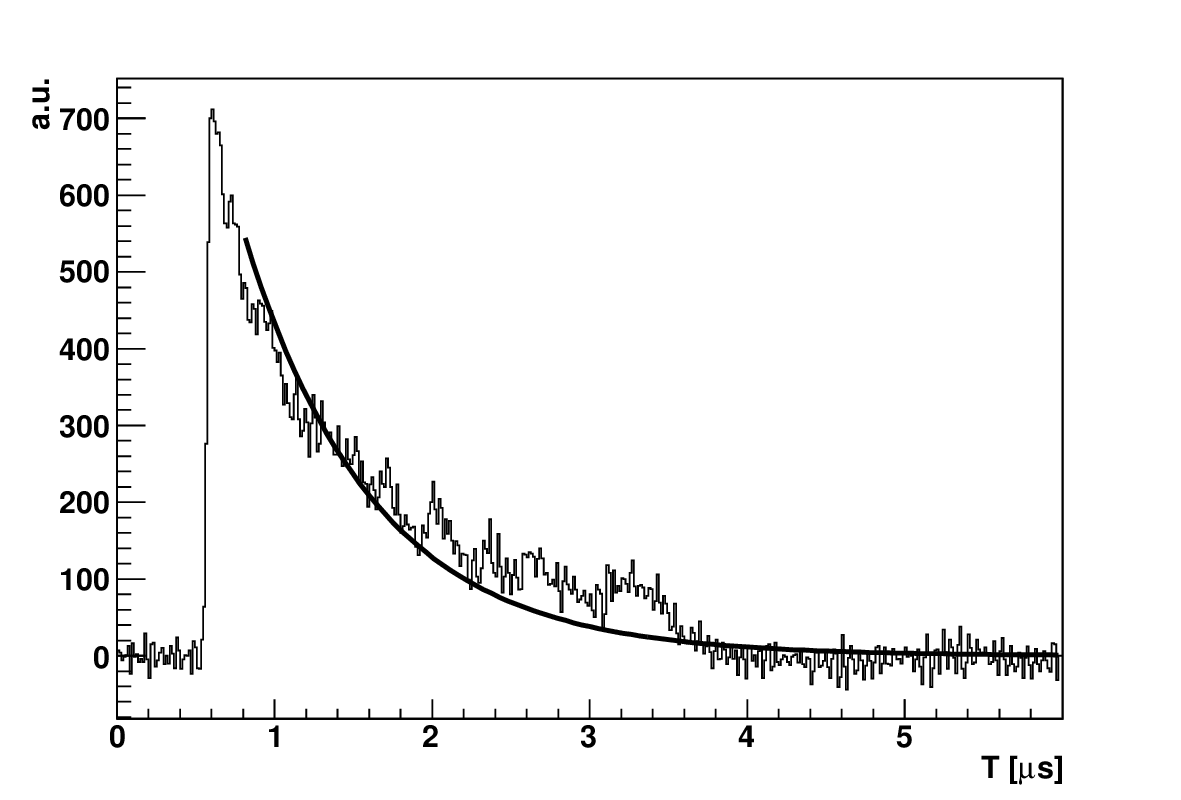}
\caption{\label{fig:pulse_shape}Pulse shape recorded with one SiPM and the sum of twelve channels from the same event. 
In this event 1495 SiPM pixels fired in total out of which 128 in the channel shown on the left.  
The event is probably produced by a cosmic muon. The fitted decay time of the pulse is 823 ns as seen by the exponential curve.
The fast component of the scintillation light is not resolved by the slow electronics and the 75 MHz DAQ. 
The shown waveforms are obtained with differentiation from the pulses recorded with the charge sensitive amplifiers.}
\end{figure}

Because more than one photon can hit the same pixel simultaneously the SiPM is an inherently nonlinear device.
The deviation from the linear response is negligible when the number of p.e.'s is small compared to the number of pixels. 
Using the estimate in Sec.\ref{Yield_estimate} we expect about 1000 pixels to fire for a 2.6 MeV $\gamma$. 
With 4800 pixels in total the deviation from linear is only a few percent.
Furthermore the hits are distributed within a time window of about 6 $\mu$s so we can safely assume that the SiPMs 
operated always in the linear range and correction factors were not applied. Fig.\ref{fig:pulse_shape} shows 
a typical high energy event with 1495 pixels fired. 
The single channel signal amplitude (Fig.\ref{fig:pulse_shape} left) does not exceed 50 pixels.

\begin{figure}[h!]
\centering
\includegraphics[width=0.9\columnwidth]{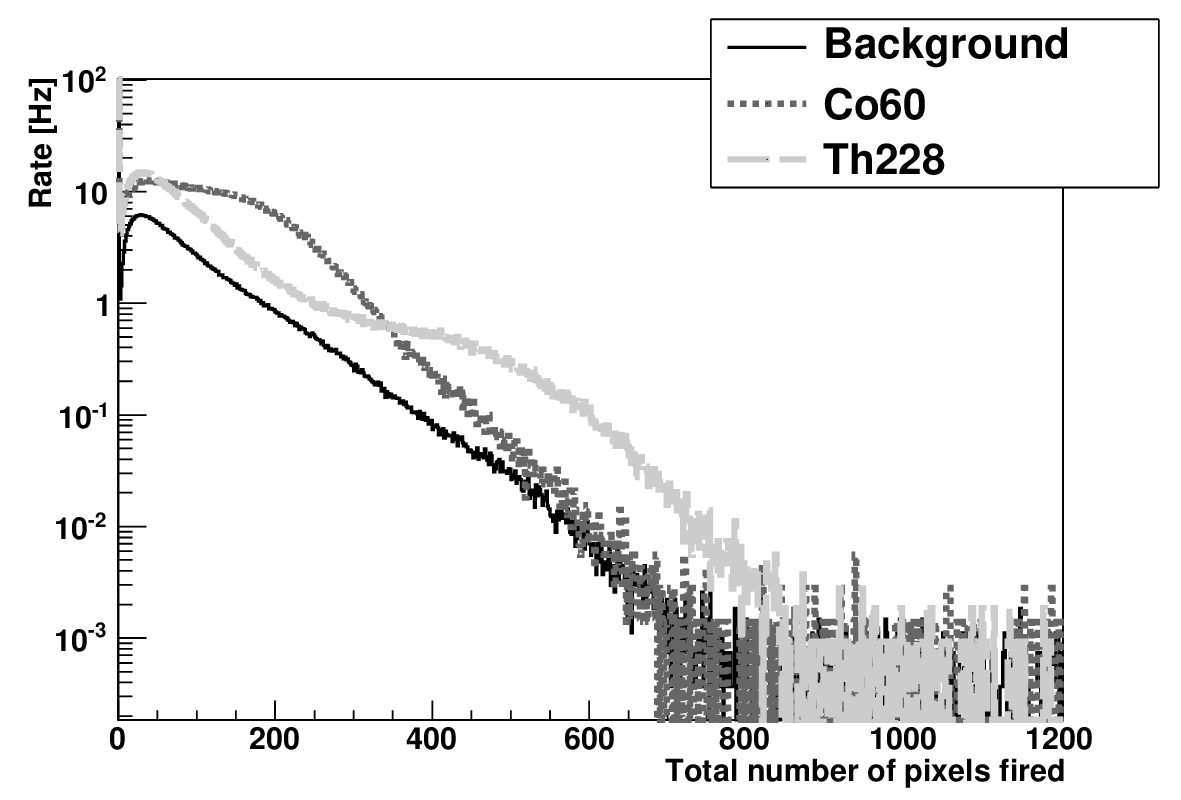}
\caption{\label{fig:ligth_collection} Response of the detector to the $^{60}$Co and $^{228}$Th sources}
\end{figure}

The spectra recorded with $^{60}$Co and $^{228}$Th sources show clear differences (Fig.\ref{fig:ligth_collection}). 
After corrections for background, cross-talk and after-pulses,
Gaussian fits to the features seen in the histograms recorded with
the $^{60}$Co and $^{228}$Th sources (Fig.\ref{fig:ligth_collection}) yielded 
about 100  p.e./MeV.

The reduced light yield can be largely explained with the quenching of the scintillation light by impurities in the argon. 
It is well known that trace elements in ppm concentration can dramatically reduce the light yield. 
The argon used was not the highest purity (Westfalen 4.6) and using a glass dewar we could not evacuate our setup before filling with LAr.

The Ar scintillation light has two components:
a short component coming from the singlet excited molecular state $^1\Sigma^+_u$ 
with 6 ns decay time and a long component from the triplet state $^3\Sigma^+_u$ with 1.6 $\mu$s decay time \cite{Hitachi}.
The quenching affects mainly the long component. 
From the fit in Fig.\ref{fig:pulse_shape} (right) one can see that the triplet decay time in our experiment was only about 800 ns.

The ratio of the intensities from the singlet to the triplet state is 0.3 according to \cite{Hitachi}. 
The measured triplet life time implies that only 60\% of the maximum  scintillation light is produced. 
Including the losses in the optical coupling, our estimate (Sec.\ref{Yield_estimate}) is therefore reduced to about 160 p.e/MeV, 
comparable to the result of the fit within the uncertainties.  

\section{Anti-Compton veto}
\label{section4}
Despite the reduced light yield our system could still perform reasonably well as an anti-Compton veto for a HPGe detector. 
We demonstrated this by inserting a HPGe detector in our setup. 
The HPGe detector was a p-type six fold segmented detector made by DSG Detector Systems GmbH. 
It had a leakage current of about 6 nA and the best resolution 
achieved was 15 keV FWHM at 1332.5 keV on the core channel.

The DAQ was triggered on the HPGe detector. Pulse shapes of the HPGe detector for the core channel and all segments 
plus the pulse shapes of the twelve SiPMs were recorded simultaneously.
As in the previous experiment we recorded 6  $\mu$s long pulse shapes for each channel.
Data was taken with an external $^{228}$Th source.

\begin{figure}[h!]
\centering
\includegraphics[width=0.9\columnwidth]{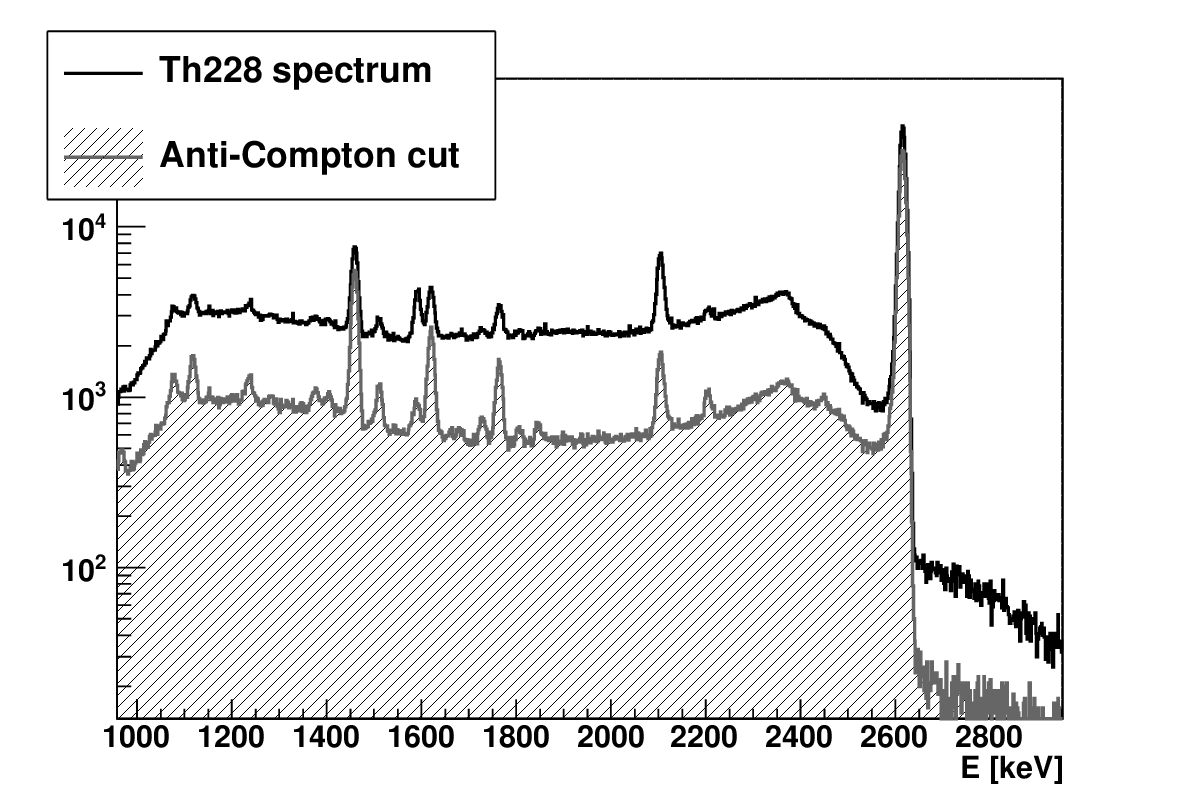}
\caption{\label{fig:anti_Compt_log} $^{228}$Th spectrum recorded with the HPGe detector. 
The dashed histogram is the same spectrum after the anti-Compton cut is applied with 0.8 p.e. threshold.
To reduce the data quantity the trigger threshold was set to 1 MeV.}
\end{figure}

The anti-Compton veto was applied off-line.
From the recorded spectrum we produced a second spectrum by removing 
all the events when at least one SiPM registered a hit within  
a coincidence window of 6$\mu$s with the HPGe signal.
Figure \ref{fig:anti_Compt_log} shows the $^{228}$Th spectrum recorded with the germanium detector and the suppressed spectrum. 
In the Region of Interest (ROI) of the GERDA experiment (2039 keV $\pm$ 50 keV) the 
flat (Compton) background was suppressed by a factor 4.2. In other words about 76\% of the background events were removed.  

\begin{figure}[h!]
\centering
\includegraphics[width=0.9\columnwidth]{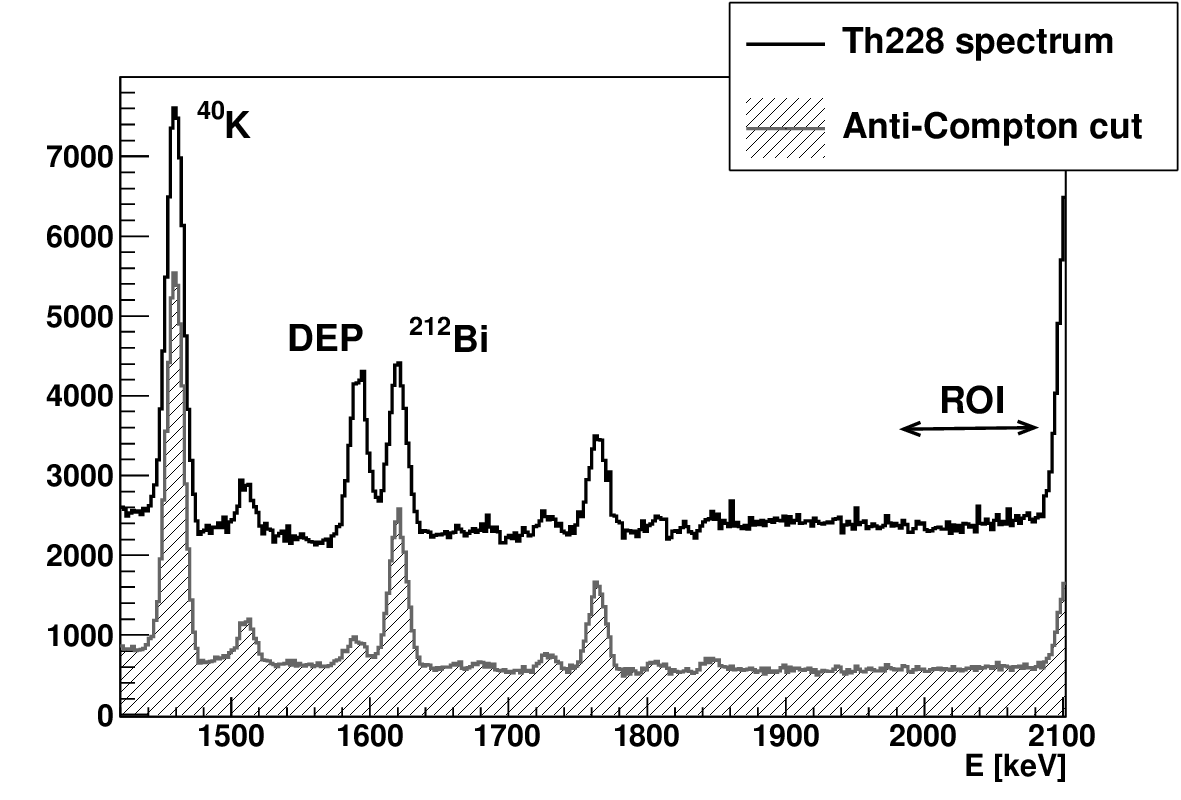}
\caption{\label{fig:anti_DEP}Suppressed spectrum of $^{228}$Th. One can see that the DEP is suppressed more than the neighboring gamma lines.}
\end{figure}

The Double Escape Peak (DEP) of the 2.6 MeV line allows further study of the Compton veto capabilities of our system (Fig.\ref{fig:anti_DEP}). 
After an $e^+ e^-$ pair production by a $\gamma$, the positron annihilates at rest and the two 511 keV $\gamma$ photons escape the HPGe detector.
The energy deposited in the HPGe detector is the initial $\gamma$ energy minus 1022 keV.
The two 511 keV $\gamma$s can deposit all their energy within a short distance in LAr.
The net effect is that up to 1 MeV can be converted to scintillation light in the same time when 1592 keV energy is deposited in the HPGe detector.
These events should be vetoed with high probability. On the other hand $\gamma$ photons fully absorbed in the HPGe detector should not be vetoed.

As we can see in Fig.\ref{fig:anti_DEP} while 
the DEP peak is suppressed by a factor 6.2 the neighboring gamma line (1620 keV from $^{212}$Bi ) is only suppressed by a factor 1.16 with 0.8 p.e. threshold. 
The strong suppression of the DEP proves that the reduction of the Compton background is not due to a random coincidence.

The $^{40}$K line was suppressed by a factor 1.1. Since the 1460 keV $\gamma$ from the $^{40}$K decay does not come in coincidence 
with any other $\gamma$, its suppression is a measure of signal acceptance. 
The 20 kHz count rate from the SiPMs in the presence of the $^{228}$Th source 
and the long 6 $\mu$s coincidence window explains the slight suppression of the $^{40}$K line.

\section{Discussion}

In \cite{LArGe} a suppression factor of 17 is reported around 2 MeV which was achieved with 
a photo-electron yield of 410 p.e./MeV. 
Our results show a lower suppression which is not only because of the reduced photo-electron yield.
Another reason is that our source was located outside the LAr volume. 
We could not introduce a source without letting air inside the dewar and contaminating the argon. 
On the other hand $^{208}$Tl always decays through a cascade and there are always two or more $\gamma$ photons emitted in the 
same time. By consequence if the source is inside the LAr volume an event can be vetoed 
even without detecting the Compton scattered $\gamma$ that exits the HPGe detector.

In order to estimate the effect of the source position we simulated our setup with Geant4. 
In the simulation we implemented a LAr volume with the same dimensions as our dewar with a germanium cylinder in the middle
corresponding to our HPGe detector (see Fig.\ref{fig:wls_setup_1}).  
We ran simulations with a pointlike $^{208}$Tl source inside and outside of the LAr volume. 
The source was positioned at the same height as the HPGe detector, as in the real experiment.   
Only the total energy deposited in the LAr and germanium volumes were saved and used for analysis.
We tuned our threshold cut on the energy deposited in the LAr to match the suppression factor obtained in the real experiment.
This was achieved at 130 keV. 
With the same threshold, a suppression factor of 8.3 could be achieved 
around 2 MeV when the source is inside the dewar, about a factor two less than achieved with the PMT setup in \cite{LArGe}. 

Our Monte Carlo studies indicate that our effective threshold on the energy deposited in LAr is about 130 keV.
Originally in the GERDA proposal \cite{GERDA2} a detection threshold for the energy deposited in the LAr of 100 keV was foreseen.
We think the goal can be achieved by improving the quality of the argon. Other improvements in the light collection should allow 
for thresholds well below 100 keV.  

We have demonstrated that the combination of SiPMs with wavelength shifting fibers 
is a viable alternative of large area PMTs for the detection of light in liquid scintillator experiments. 

The full potential of the anti-Compton veto based on SiPMs and WLS fibers is not reached yet. 
Our setup suffered mainly from the quality of the argon.
In Section \ref{Yield_estimate} we showed that
the effect of the attenuation in the fiber can be minimized by using many short fibers. 
Increasing the number of fibers while keeping the number of electronic channels low would be possible in combination 
with larger area SiPMs that are already on sale. 
With further optimization the limit imposed by the trapping efficiency of the fibers should be 
possible to achieve.

The setup described in this paper was built with regard to the needs of double beta decay and dark matter experiments.

We therefore conclude that SiPMs represent an attractive solution for background reduction in low background experiments.

\section{Acknowledgment}
We would like to thank Matthias Laubenstein for the screening measurements. The measurements were performed with the GeMPI detector
based on the developments originating from the work done at the Max-Planck-Institut fuer Kernphysik (MPIK)
in Heidelberg (Germany). The detectors are currently operated, maintained at the LNGS.

This work was partly supported by the DFG grant SFB/Transregio 27 'Neutrinos and Beyond'.

\bibliographystyle{model1-num-names}
\bibliography{SiPM_article}

\end{document}